\documentclass{article}
\usepackage{arxiv}

\usepackage[utf8]{inputenc}

\usepackage{graphicx}
\usepackage{url}
\usepackage[super]{nth}
\usepackage{siunitx}

\usepackage{booktabs}
\usepackage{subcaption}
\usepackage{color}

\title{BlockColdChain: Vaccine Cold Chain Blockchain}

\author{
  Ronan D. Mendonça\\
  Computer Science Department\\ Universidade Federal de Viçosa (UFV)\\ Florestal -- Brazil \\
  \texttt{ronan.dutra@ufv.br} \\
   \And
 Otávio S. Gomes\\
  Computer Science Department\\ Universidade Federal de Viçosa (UFV)\\ Florestal -- Brazil \\
  \texttt{otavio.s.gomes@ufv.br} \\
  \AND
  Luiz F. M. Vieira \\
  Computer Science Department\\ Universidade Federal de Minas Gerais (UFMG)\\ Belo Horizonte -- Brazil \\
  \texttt{lfvieira@dcc.ufmg.br} \\
  \And
  Marcos A. M. Vieira \\
  Computer Science Department\\ Universidade Federal de Minas Gerais (UFMG)\\ Belo Horizonte -- Brazil \\
  \texttt{mmvieira@dcc.ufmg.br} \\
  \And
  Alex B. Vieira \\
  Computer Science Department\\ Universidade Federal de Juiz de Fora (UFJF)\\ Juiz de Fora -- Brazil \\
  \texttt{alex.borges@ufjf.edu.br} \\
  \And
  José A. M. Nacif \\
  Computer Science Department\\ Universidade Federal de Viçosa (UFV)\\ Florestal -- Brazil \\
  \texttt{jnacif@ufv.br} \\
}

\begin{document}
\maketitle

\begin{abstract}
In this paper, we propose a blockchain-based cold chain technology for vaccine cooling track. The COVID-19 pandemic has caused the death of millions of people. An important step towards ending the pandemic is vaccination. Vaccines must be kept under control temperature during the whole process, from fabrication to the hands of the health professionals who will immunize the population. However, there are numerous reports of vaccine loss due to temperature variations, and, currently, people getting vaccinated have no control if their vaccine was kept safe. Blockchain is a technology solution that can provide public and verifiable records. We review the World Health Organization (WHO) cool chain and Blockchain technology. Moreover, we describe current IoT temperature monitoring devices and propose Blockcoldchain to track vaccine cold chain using blockchain, thus proving an unalterable vaccine temperature history. Our experimental results using smart contracts demonstrate the system's feasibility.
\end{abstract}


\section{Introduction}
\label{sec:introduction}

The COVID-19 pandemic has quickly affected the world and changed our lives. Unfortunately, this virus is already responsible for millions of deaths around the globe.  A solution to end this pandemic is vaccination. However, vaccines must be under the rigid control of temperature from fabrication to the hands of the health professionals who will immunize the population. Vaccines may quickly lose their effectiveness if they become too hot or too cold at any time, especially during transportation and storage. 

There are several reports in the news of vaccines lost due to temperature issues. For example, on January \nth{20}, 2021, authorities in Michigan announced that nearly 12,000 doses of the Moderna Covid-19 vaccine had been ruined by temperature control malfunctioning  during shipment, and in Maine, more than 16,000 vaccine doses were spoiled.\footnote{\url{https://www.washingtonpost.com/nation/2021/01/20/moderna-vaccine-spoiled-maine-michigan/}} 
In Boston, 1,900 Covid vaccine doses were ruined due to a loose freezer plug.\footnote{https://www.nbcnews.com/news/us-news/1-900-covid-vaccine-doses-ruined-loose-freezer-plug-massachusetts-n1255300}  
In Milwaukee, a hospital employee ruined more than 500 doses by removing vaccines from a refrigerator. 
\footnote{\url{https://www.usatoday.com/story/news/nation/2020/12/31/wisconsin-covid-19-vaccine-ruined-aurora-employee/4097416001/}}
In the city of Rio de Janeiro, hundreds of COVID-19 vaccine shots ---that should be kept between \SI{2}{\celsius} and \SI{8}{\celsius}--- were wasted after a power outage.\footnote{\url{https://www.usnews.com/news/world/articles/2021-01-29/rio-may-throw-out-hundreds-of-covid-19-vaccine-shots-after-power-outage}}

The \textit{cold chain} is a term used to refer to the supply chain in which goods, such as vaccines, are maintained in a temperature-controlled environment. The World Health Organization (WHO) has guidance for the vaccine cold chain~\cite{WHOColdChain}. It involves monitoring vaccine temperatures from the point of manufacture until administration in a health center.

Despite the notable importance of vaccines' cold chain, end-users have no access to information, including vaccine conditions and temperature along the vaccine supply chain. Often, there is no transparency and no routine monitoring in the cold chain system. Moreover, due to the lack of accountability, vaccines may get lost or stolen in any part of the supply chain. These issues contribute to the loss of about 30\% of vaccines during transportation \cite{hasanat2020iot}.

Vaccine manufacturers and distributors may use IoT technology to collect the cold chain pieces of information. In this way, temperature and humidity sensors monitor the cold chain, and special communication units send necessary notifications and text messages to the healthcare supervisors accordingly. The entire monitoring process is carried out automatically, leaving no space for human errors or negligence \cite{hasanat2020iot}. However, data are still stored in centralized databases, belonging to private entities despite the automatization process. These data can be easily forged along the supply chain process.

Blockchain is a recent disruptive technology. It is a form of immutable distributed ledger, which allows one to transfer ownership, record transactions, and track assets without the need of third parties.
Considering the properties of blockchain technology and the importance of the vaccine cold chain, in this work, we present BlockColdChain: a Vaccine Cold Chain Blockchain-based solution to aid the fight against the COVID-19 pandemic by providing an IoT-based intelligent approach to verify and audit the vaccine cold chain.  BlockColdChain receives all cold chain steps sensor data and securely stores them in a blockchain. Consequently, BlockColdChain gives transparency to the routine of monitoring the cold chain system and accessing information, which may also reduce the losses, the tampering, and the stealing of vaccines.

We organize the remainder of this paper as follows.
Section~\ref{sec:background} introduces the background. Section~\ref{sec:related} presents the related work. Section~\ref{sec:IoT} overviews IoT temperature monitoring devices. Section~\ref{sec:Blockcoldchain} details the blockchain for the vaccine cold chain. Section~\ref{sec:analysis} describes the solution analysis. Finally, Section~\ref{sec:conclusion} concludes this paper.

\section{Background}
\label{sec:background}

In this section, we briefly overview vaccines cold chain, mostly supported by the WHO definition. Then, we introduce blockchain technology.

\subsection{Cold Chain}

As stated by the World Health Organization (WHO): ``Vaccination is a simple, safe, and effective way of protecting people against harmful diseases, before they come into contact with them.'' However, vaccines are extremely sensitive to temperatures and must be stored within a specific safe temperature range. For example, for some vaccines, WHO has fixed the temperature range for vaccine storage and transportation as +2 to +8°C, and vaccines completely lose potency if they are exposed to
temperatures beyond this range even for short durations. 
Some specific vaccines require temperatures as low as -80°C, requiring specialized monitoring solutions.

\begin{figure}[!ht]
\centering
\includegraphics[width=0.99\columnwidth]{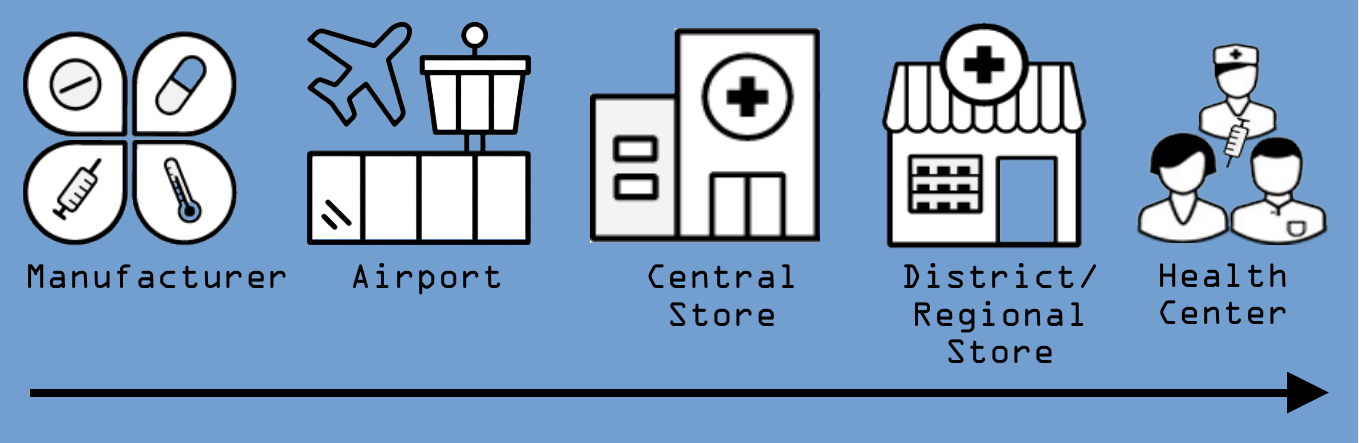}
\caption{Cold Chain path.}
\label{fig:coldchain}
\end{figure}

The uninterrupted refrigerated supply chain of vaccines, maintained in a temperature-controlled environment, is known as the cold chain. The cold chain path, defined by WHO, is shown in Figure~\ref{fig:coldchain}. At a glance, a manufacturer produces a vaccine. Then, it is transported to an airport (or any other transportation system), thus reaching a central store. Vaccines are distributed to regional stores, and finally, they arrive at a health center where they are administered.

The process of monitoring the cold chain is known as cold chain monitoring. In this work, we consider IoT-based systems to monitor the cold chain. In this sense, data loggers, temperature and humidity sensors, and low-power transmission devices are used to perform cold chain monitoring.

Different types of equipment are used in a cold chain.
At a national level, depending on the capacity required, vaccine storage and transportation usually require cold or freezer rooms, freezers, refrigerators, cold boxes, and, in some cases, refrigerated trucks for transportation. At an intermediate level, such as district/regional, the vaccine cold chain usually employs cold and freezer rooms, freezers, refrigerators, and cold boxes. In some cases, at this level, refrigerated trucks are used for transportation. Finally, the health center facility usually uses refrigerators (in certain instances with water pack freezing/cooling compartments), cold boxes, and vaccine carriers to maintain all vaccines.

\subsection{Blockchain}

Blockchain is considered a distributed technology of data sharing, composed of countless participants in which it is not necessary to exist a centralized control. The use of blockchain technology is numerous, ranging from validation, integrity to interoperability of data~\cite{GORDON2018224}. A blockchain can offer more transparency and security to transactions~\cite{francisco2018supply}, which is a demand of a cold chain.

\begin{figure*}[!ht]
\centering
\includegraphics[width=0.9\textwidth]{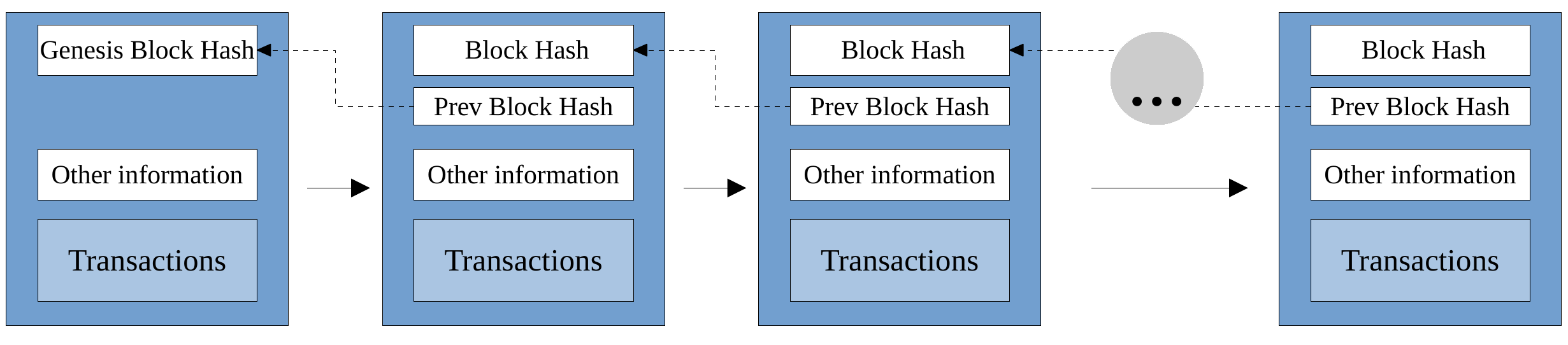}
\caption{Block structure.}
\label{fig:blocos}
\end{figure*}

The blockchain is a growing list of blocks linked using cryptography. Each block contains a hash of the previous block, timestamp, and transaction data. It works like a distributed ledger, organized in a chain of blocks through the link of encrypted hashes. More in deep, Figure~\ref{fig:blocos} presents an overview of a blockchain. In this example, each block contains a group of transactions. In a general form, transactions may be a transfer of ownership (e.g., money exchange) or a record registration, such as a database entrance. Moreover, blocks contain the previous block hash. A block may also include some general information. Finally, a block contains a hash signature of all its content. Note that, as the previous block hash is part of the current block content, the hash of the current block also depends on the content of the previous block. As a consequence, the current block is linked to the previous block. One cannot change past blocks without changing the whole following sequence of blocks.

Blockchains do not require trust in a third party or costly centralized institutions to execute transactions.  Their execution follows a distributed consensus mechanism, where all system users may participate.
In most blockchain-based applications, there are two main ways to reach a consensus: the Proof of Work, a protocol based on a user effort, and the Proof of Stake, based on the users' currency stake~\cite{de2020analysis}.  

Blockchains may organize their users and special participants to be more effective. In this sense, there are three main types of blockchains: public, consortium, and private. Each kind of blockchain organization contains specific characteristics that must be taken into consideration for its adoption. For example, public blockchains are decentralized, permissionless, and their data are public. On the other hand, private blockchains are centralized, permissioned and their data can be public or restricted. By contrast, consortium blockchains are partially centralized, permissioned, and their data can be public or restricted. 

The consortium blockchain is considered the most indicated for applications that involve participants in different organizations that may have egalitarian control over the blockchain network. In a consortium-based blockchain, one can set different permission configurations for the network participants, thus restricting each participant to the previously authorized data (or transaction data). Moreover, one can set a distinct type of participants as specific participants to deal with the consensus mechanism of the network.

New technologies have emerged to improve the functioning of the blockchains. This is the case of Ethereum,\footnote{\url{https://ethereum.org} -- last accessed 02/27/2021} which has the execution of \textit{smart contracts} in its architecture. These contracts are algorithms~\cite{buterin2013ethereum} inserted in this network, and that can be sent by blockchain, have automated behaviors, do not require third parties, and provide greater security to all parties.

\section{Related Work}
\label{sec:related}

Here we discuss other papers related to blockchain technology usage in the supply chain and vaccines' traceability. As blockchain technology shows benefits of its usage in various segments, it is becoming more suggested and accepted in areas that require security and disposal of information to unknowns. Blockchain can be used in the vaccine distribution chain to store and share data so that its authenticity can be easily verified. According to~\cite{kshetri20181}, blockchain offers transparency through the immutability of transactions, increasing confidence for the concerned parties and providing traceability.

In the paper presented by~\cite{kamble2020modeling}, the authors identify thirteen facilitators and establish the relationships between them for the adoption of blockchain in the supply chain. The facilitators were validated by specialists and were submitted to a methodology combining Structural Interpretative Modeling with Evaluative Laboratory and Decision Making Testing to visualize the relationships between the facilitators and the adoption of the blockchain technology. The bibliography shows that blockchain technology offers various benefits, leading to improvements in the sustainability performance of supply chains and the elimination of problems related to confidence in the chain. This study suggests that, among the identified facilitators, traceability was the main reason for the implementation of blockchain, followed by auditability, immutability, and origin. They also conclude that it is essential to adopt the technology to take advantage of its various benefits.

The authors of the paper~\cite{kamilaris2019rise}, researched the impact of blockchain technology in agriculture and the supply chain. They present projects and initiatives being currently developed and discuss general implications of these projects and indicate traceability as an opportunity and a potential benefit. Among the positive aspects listed by the authors, we can mention the reduction of transaction cost, reduction in intermediaries dependency, higher transparency, and fewer frauds. Though, they also indicate difficulty in accessing the technology as the main barrier for its adoption.

The following papers used a case study for implementing blockchain technology in the context of traceability. In~\cite{bumblauskas2020blockchain}, the authors propose an implementation for the traceability of production. The objective of the paper is to verify the precision and transparency in applying blockchain in the traceability of the supply chain to give consumers information about the items being bought by them. For the implementation, the authors used IoT sensors, the Hyperledger blockchain, and a web application to provide an infrastructure capable of capturing data and analyzing the impact of its implementation. The authors conclude by stating that blockchain technology complements the traceability of the supply chain in a setting that already holds solutions that attend part of the concerned parties. They also note that the changes required for blockchain technology adoption do not significantly affect already existing processes. The technology's potential is revolutionary for the supply chain in the scope of transparency of the process to the final consumer. 

OriginChain~\cite{xu2019designing} was developed to provide traceability blockchain-based. For the development of the application, the authors based the requirements in a traceability system of a partner and designed their architecture adapting to the conventional already existing database. To provide immutable traceability of data, the authors tested the application using real data of users and products. To develop the infrastructure, they used an Ethereum based blockchain in a private and public manner. The authors evaluated the application quantitatively through reading and writing latency in the blockchain, a local database, and a remote database. The discussion about the results indicates the performance and privacy as limitations to be considered and suggests that private blockchains have better performance than public ones.
 
The paper presented by~\cite{figorilli2018blockchain} implements an architecture using blockchain for the traceability of the supply chain. The authors simulate the traceability system to gather data about the supply chain that uses RFID sensors. They used the Workbench Azure Blockchain as infrastructure for the development of the blockchain application. According to the authors, the blockchain technology used for the traceability of products is economically viable and guarantees reliability, transparency, and security. 

Unlike the other papers, our solution uses blockchain technology to promote smooth interoperability among IoT devices and data availability to the concerned parties.

\section{Temperature monitoring devices}
\label{sec:IoT}

It is essential to monitor and record the temperature of vaccines throughout the cold chain. However, current technology does not make this information public, nor no one knows if it has been tampered with.

A vaccine vial monitor (VVM) is a chemical indicator label attached to the vaccine container. VVMs are temperature monitoring devices that accompany vaccines throughout the entire cold chain. 

Electronic temperature loggers are placed with the vaccine load in a vaccine refrigerator. They record the refrigerator temperature, usually at no more than 10-minute intervals. They record the temperature history for the last 30 days. Alarms are triggered if the temperature of the refrigerator rises above the recommended range.

Nowadays, most temperature monitoring devices are not integrated into the Internet.

\subsection{IoT Temperature monitoring devices}

The vaccine quality in the cold chain is affected mostly by its temperature. Adequate temperature can extend its lifespan, and its monitoring can transmit and resume all data collected. This requires identifying the products and constructing an information system that includes the collection and availability of data from its production until administration in patients.

Cold chain management requires fast decision-making to keep the quality and accomplish fault identification. Once the vaccines are transported through many places till they get to the vaccination center, a well-established cold chain logistics requires monitoring and automatic control of all its operations.

Implementing an intelligent and interconnected system is not a simple task since there are a lot of challenges to be solved, related to the number and quantity of sensors necessary, its placement on cargo, the reliability in the data reading, the volume of data read, and data loss during transmission~\cite{BADIAMELIS2018170}.

There are many solutions for monitoring the temperature. Some of these solutions are Time-Temperature Indicators (TTIs), Thermistors, Resistance Temperature Detectors (RTDs), and Thermocouples. Though, they present difficulty in accessing and sharing the monitored data. The passive RFID tags can be used to trace the cold chain with low cost and long durability. The RFID allows monitoring so that data can be accessed through Bluetooth communication and identification through radio-frequency~\cite{Vivaldi2020}.

According to~\cite{BADIAMELIS2018170}, Radio Frequency Identification (RFID) and Wireless Sensor Networks (WSN) are the most used technologies in cold chain monitoring. They provide more efficiency for the cold chain and reduce labor costs. The RFID can be active, passive, or semi-passive. A passive RFID, which we consider to be more effective, cheaper, and easier to use in the cold chain, sends data by reflecting the electromagnetic field emitted by a reader. It can record the temperature and its location and sync this data to a database when read. Though, both RFID and WSN have a high cost associated and, because of that, these sensing units are usually limited.

\section{Blockcoldchain}
\label{sec:Blockcoldchain}

\subsection{Architecture}

\label{sub:Architecture}

The technology used by distributed systems allows the integration of many solutions securely and effectively. Through peer-to-peer communication, for example, we can implement system interoperability. The consensus mechanism, on the other hand, weights reliability in the transactions between the related parties. Regarding distributed and encrypted storage, we secure a high level of security and availability in the network. Based on these concepts and the processes in the cold chain path, we define an architecture model to assist the transport control and storage of vaccines dealing with traceability, surveillance, storage, and temperature control.

\begin{figure}[!ht]
\centering
\includegraphics[width=0.99\columnwidth]{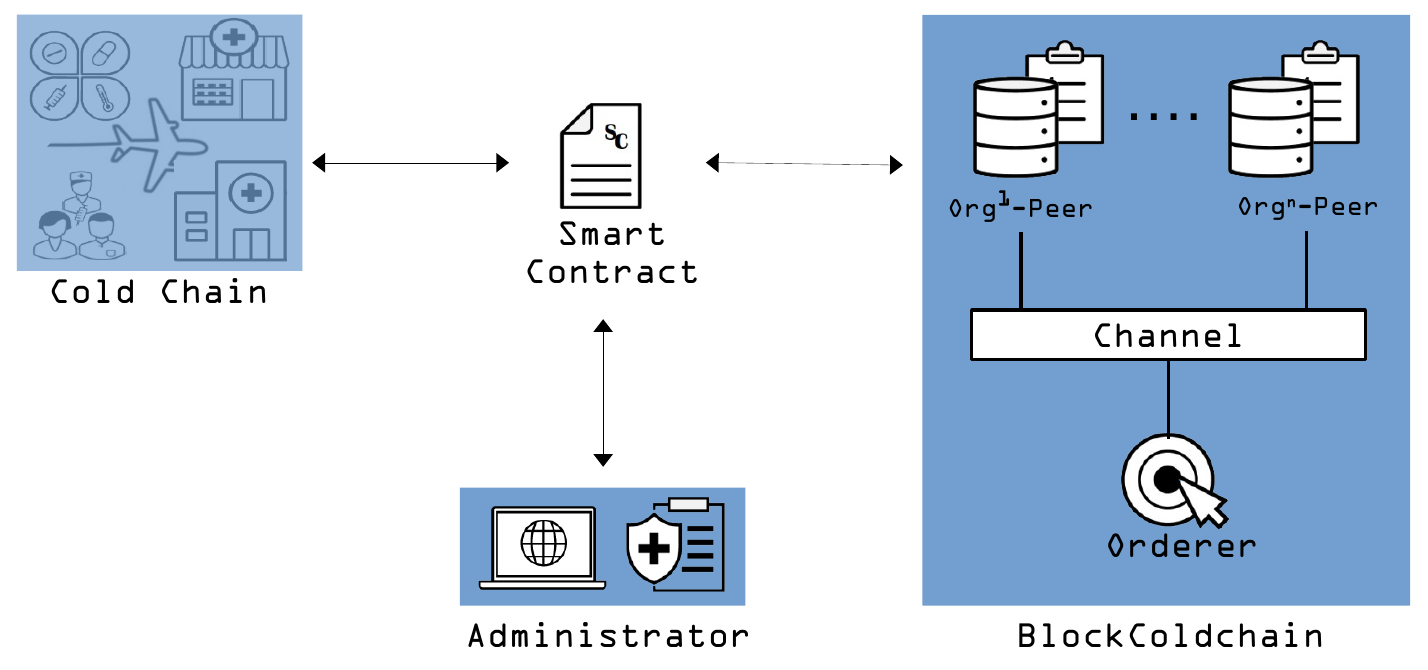}
\caption{Architecture model.}
\label{fig:Architecture}
\end{figure}

The architecture that we present aims to interconnect various components involved in the cold chain management system to guarantee that the users can rely on the data collected during the vaccine transport and storage process. In this article, one of the main reasons we used blockchain technology is to provide traceability and facilitate interoperability among the used components. Figure \ref{fig:Architecture} shows the architectural model designed to implement blockchain technology through the vaccine cold chain.

\begin{figure}[!ht]
\centering
\includegraphics[width=0.99\columnwidth]{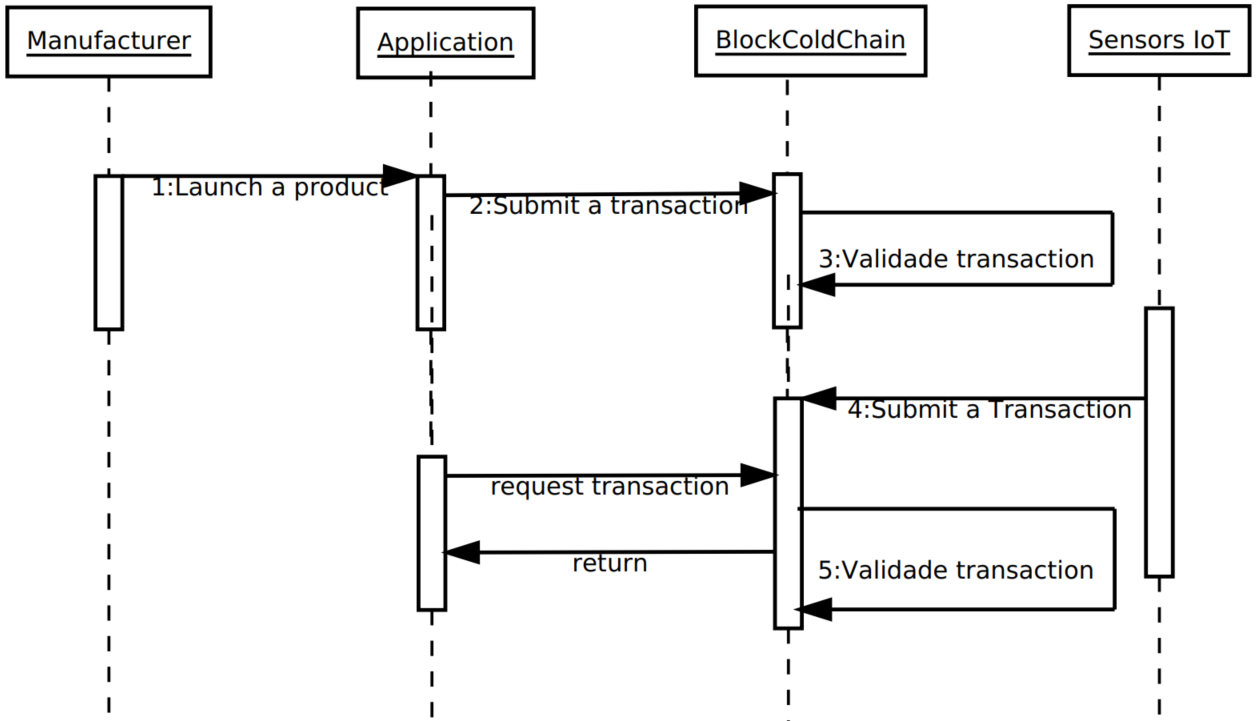}
\caption{Sequence Diagram.}
\label{fig:Diagram}
\end{figure}

The Blockcoldchain will be triggered through the whole vaccine manufacturing, storage, and transport process. Figure \ref{fig:Diagram} presents a sequence of operations in the network in which the vaccine produced by the Manufacturer must be inserted as a new item. In the steps where the vaccine is transported and stored, IoT sensors can read data and submit it to the network. The Airport, Central Store, District/Regional Store, and Health Center act actively in sensing and updating locations and values collected.

Different entities form the resulting network in this architectural model. The network configuration is performed according to each of these entities' necessities, and it is based on access, availability, and consensus over the transactions. The components of the Blockcoldchain, according to Figure \ref{fig:Architecture}, are the Organization types (Org), Peer, Orderer, Channel, and the Smart Contract. The organizations represent each real entity in the network, like a company. The Peers are the nodes from one Organization and can be used to propose and validate transactions in the network. On the other hand, the Orderers order the transaction in blocks and control the business rules that will define the transaction format. It is through them that the external applications and IoT sensors will send and read sensed data.

In this case, the Manufacturer requests a transaction that the product is launched (labeled (1) in Figure \ref{fig:Diagram}). The proposal is delivered to the application. The application submits the transaction to the BlockColdChain (labeled (2)). BlockColdChain executes the smart contract to validate the transaction (labeled (3)).  Each sensor submits a transaction. The BlockColdChain confirms that the transaction satisfies the conditions and then commits the transaction (labeled (4), (5)). Finally, the application requests the transaction and checks the results received from the BlockColdChain.

Using blockchain, we propose an infrastructure capable of dealing with the entire vaccine distribution chain. The main component of this infrastructure is the smart contract. This contract is responsible for controlling all the vaccine cold chain information, from its production to its administration in the population. The flow followed when deploying and using the contract works as follows. The user that deploys the contract in the blockchain is set as the contract administrator. The administrator is responsible for managing the users in the network. The administrator can add new administrators and add and remove users whenever is necessary. The administrators can be entities such as the World Health Organization (WHO) and other minor entities or even governments.

On the other hand, the users represent the places in which the vaccine can stay for a period in the cold chain. These users can represent cold rooms, freezers, refrigerators, refrigerated trucks, or any other places where vaccines can stay for a period and need refrigeration. The information required from these users is the items inside them, when they arrived, and when they left. Also, they should insert information about their temperature regularly. The users can define this regularity. The users must insert their data into the blockchain. This process can be managed by temperature monitoring devices, as mentioned before, and position monitoring devices through the whole chain.

With the information adequately recorded, the user can check a list of the items inside it anytime and the history of all the temperatures measured. Also, the administrators can, at any time, find the information about a particular product in the cold chain. It can access the users, see the product, the time it arrived, and all the measured temperatures while this product was kept in this place. This way, the administrators have full access to the data in the network and can find problems with the temperature of the vaccines in any location in the cold chain.

On top of that, any person can use the vaccine identification and the smart contract to find all the information about the vaccine, effectively tracking the places in which the vaccine has been, as well as its temperature through the whole chain. With the smart contract described, the foundation of the entire system is set.

\begin{figure}[!ht]
\centering
\includegraphics[width=0.99\columnwidth]{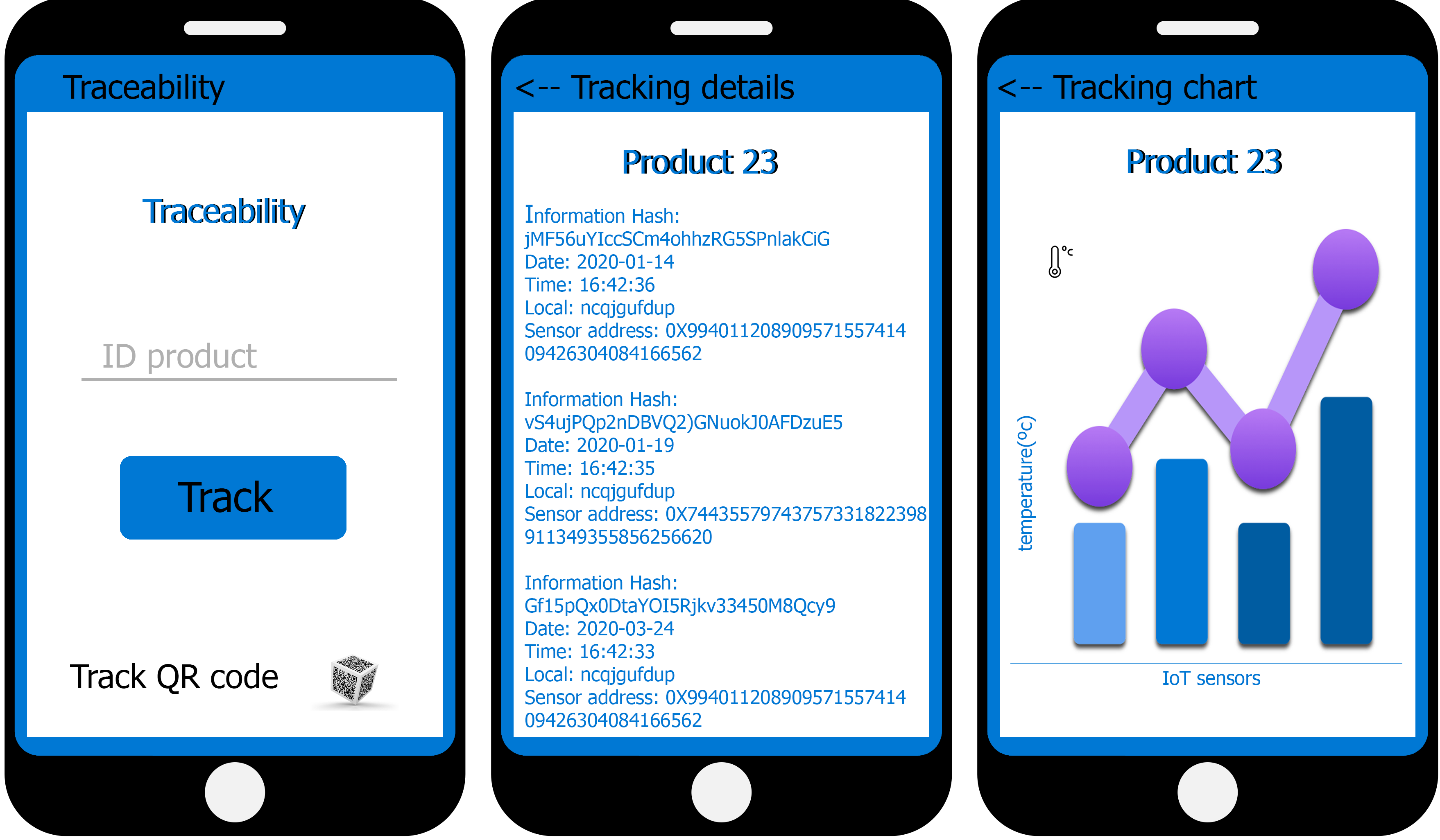}
\caption{App screens.}
\label{fig:screen}
\end{figure}

There can be multiple ways for the final user to access the information in the blockchain. There should be three different interfaces for the users to access the system. The first interface is an application for the administrators to control the entire process. This interface is a computer program in which the administrators can call any function of the contract and see the information about all users and all products in the chain. The second interface is the website in which the users will add information in the network about themselves. The last interface is the one that allows anyone to check if the vaccine they are getting was well maintained through the whole vaccine distribution chain. This can be a smartphone application that works to access the blockchain and query a request about a vaccine. It should show the final user the history of all the places in which the vaccine was and the temperature of each place while the vaccine stayed in there. The app  (Figure \ref{fig:screen}) was structured to operate during the transit of the cold chain path or monitoring phase.

\subsection{Distributed Consensus.}
\label{sub:Consensus}

The consensus works with an algorithm in a distributed way, intending to generate a new block and be accepted among the network nodes. The algorithms consist basically in a solution of a dilemma to obtain a consensus among the nodes in the network over a transaction. Among the nodes, there can be some nodes to disrupt the process. The objective is to obtain the consensus among the valid nodes and ignore any attempt of obstruction. The comparison between the consensus mechanisms follows the criteria of energetic efficacy, the entrance of new nodes, and fault tolerance.

A more profound study about these algorithms can present more details about the capacity of storage and time spent to generate the blocks in each algorithm. The choice of the consensus mechanism can have a significant impact on the performance of the intended solution. Therefore, it must be considered the application requirements concerning the storage of data and response time.

\section{Analysis}
\label{sec:analysis}

Blockchain technology was used as a decentralized database that promotes interoperability among the cold chain path and sensing entities. It can manage transactions with efficiency because it maintains data shared among every network node meaning that the information is in every network node and, therefore, can not be modified. Thus, this technology presents high transparency and efficiency, allowing the insertion of trustworthy transactions in an unreliable environment.

This work introduces blockchain technology for monitoring vaccine temperature and offers traceability from production to application. Many aspects must be discussed considering the advantages of the introduction of traceability in the cold chain. The possibility of global monitoring is primarily accomplished by IoT sensors that send the vaccine status, temperature, and current location.


We have implemented and executed experiments in the platform Ethereum in the form of a private network to store and access data safely in the cold chain. In our system, the smart contract is responsible for managing the IoT sensors, the tracked items, and the data related to each item. Then we implemented the smart contracts, an application for generating and inserting data on the network, and a mobile application for product tracking.

\begin{figure}[ht]
\begin{subfigure}{.5\textwidth}
  \centering
  \includegraphics[width=.8\linewidth]{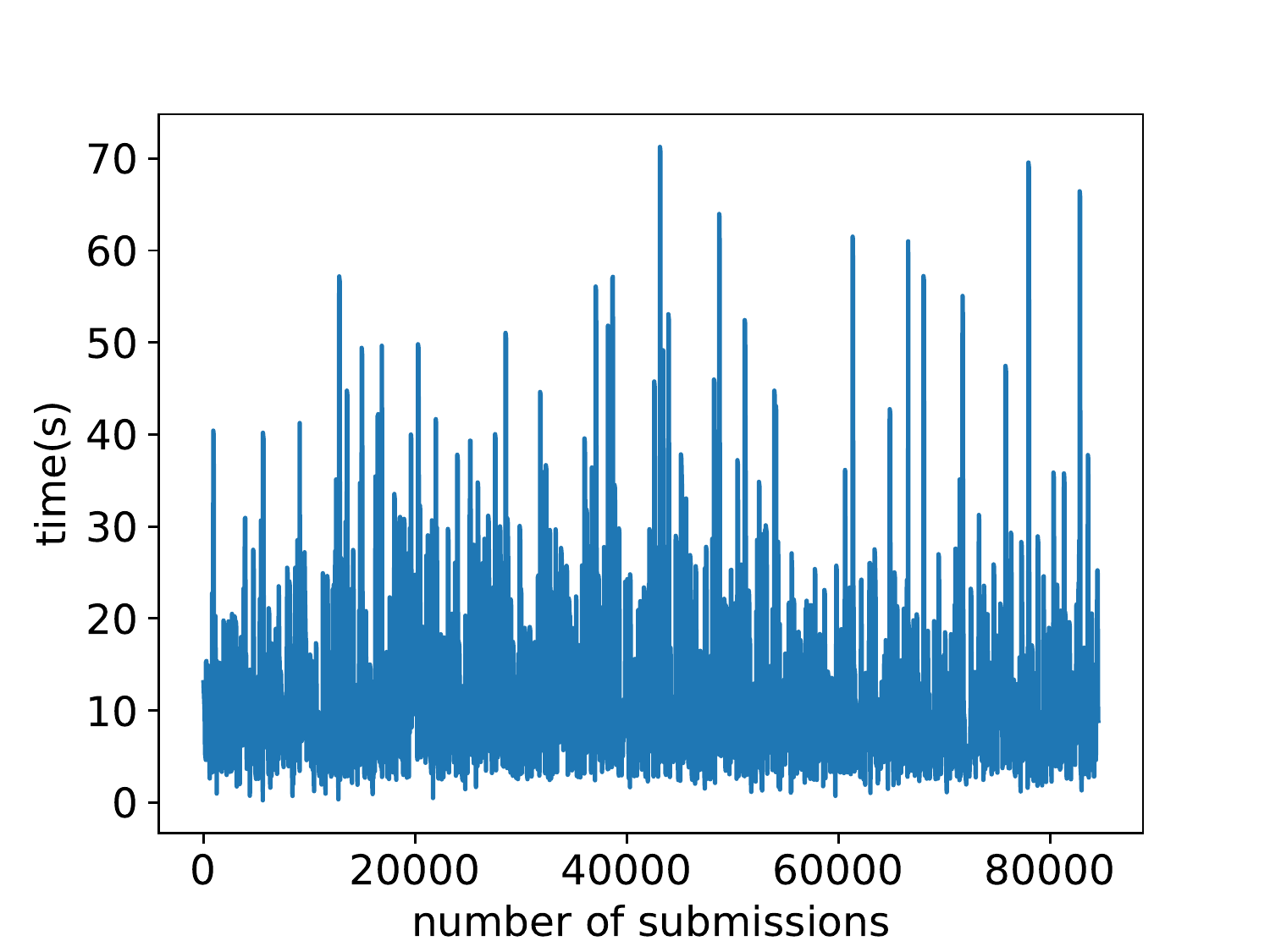}  
  \caption{Validate submit transaction}
  \label{fig:sub-first}
\end{subfigure}
\begin{subfigure}{.5\textwidth}
  \centering
  \includegraphics[width=.8\linewidth]{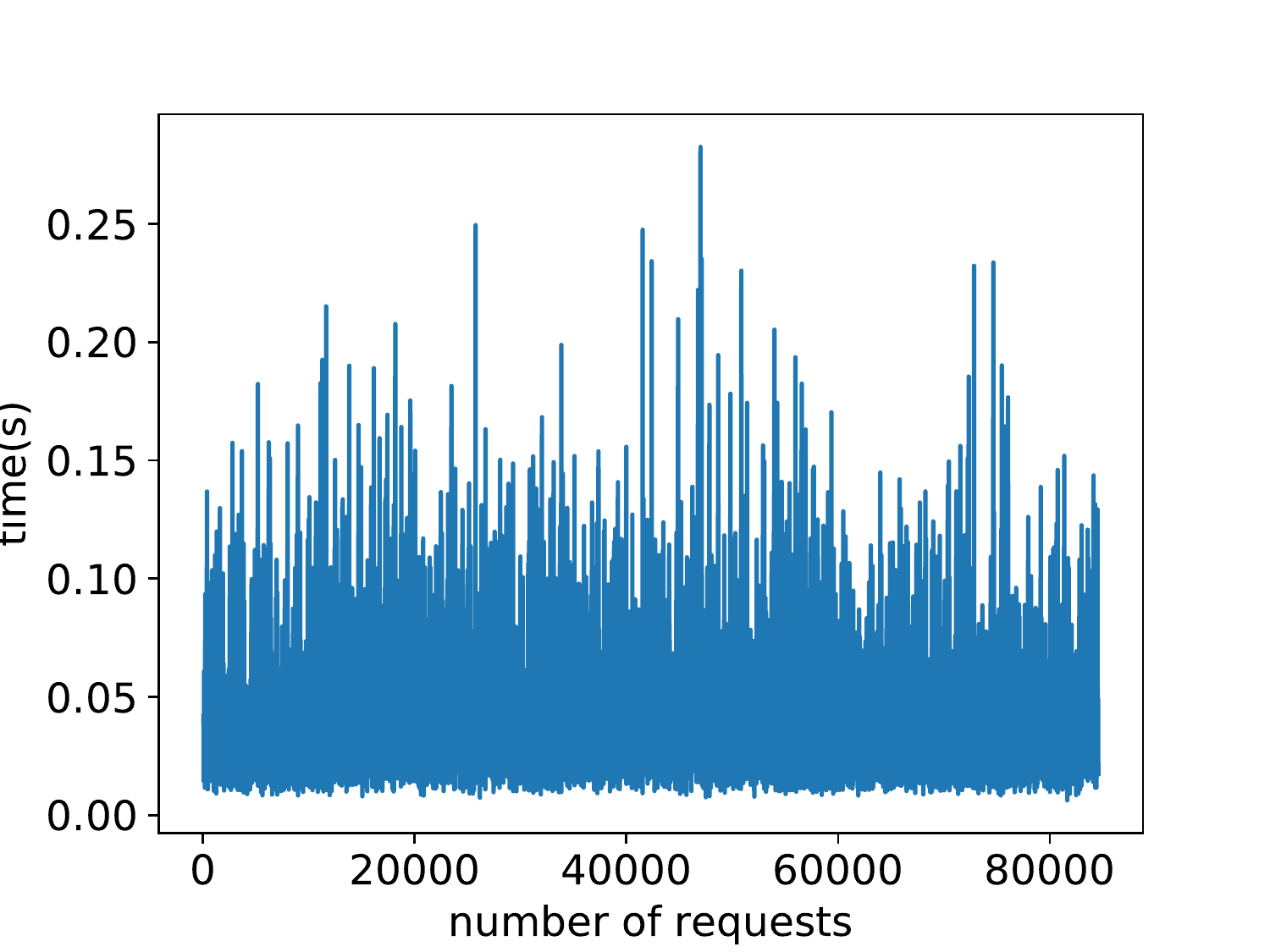}  
  \caption{View data}
  \label{fig:sub-second}
\end{subfigure}
\caption{Response time to validate a transaction and view data.}
\label{fig:graphic}
\end{figure}

Figure \ref{fig:graphic} shows the response time and the number of submissions of data reading and writing operations using smart contracts to manage access control policies. To read the tracking data, it is necessary to send, by the external application, only the product code so that the data can be returned to the user. The write operation requires the signature of the administrator or the IoT sensor plus the item data. The blockchain checks access permissions and sends the data to storage, saving only a hash of the data on the blockchain.

From Fig \ref{fig:graphic}(a) and Fig \ref{fig:graphic}(b), we can see that the time cost is greater when submitting transactions than when visualizing data. This happened because, in the blockchain system, during the view operation, the system does not need to wait for the consensus. Therefore, part of the monitoring overhead can be reduced for the cold chain. The time overhead of the blockchain system is mainly consumed in the consensus process. 

In the simulation experiments, the transactions took, on average, 11 seconds to validate a data submission transaction on the network and 0.05 seconds to return data. The longest time took to validate a transaction did not exceed 71 seconds and the data visualization took a maximum of 0.28 seconds. Therefore the time obtained perfectly satisfies the requirements of the cold chain.

A common point between the solution presented and those presented in the work of \cite{hasanat2020iot} is the use of an application to monitor the data and all actions that change the state of the IoT sensors. But the differential of the solution of this work is in the possibility of granting, or not, access to the data directly by a client application, by saving the data inside the blockchain, and by providing easier interoperability.

Another important aspect that could benefit the evaluation of the vaccine is the possibility of precisely defining the quantity that will be used and lost in the packages that will get to their final destination in the cold chain.

Furthermore, the system offers the possibility to detect flaws in equipment where vaccines are being stored since the values sensed and sent to the blockchain can be audited and traced. Also, the traceability of vaccines with the introduction of blockchain offers the opportunity to extract ``retroactive inferences" about products in each step of the cold chain path.


\section{Conclusion}
\label{sec:conclusion}

Blockchain technology has the potential of disrupting the way monitoring and traceability are conducted in the cold chain. Using blockchain can be an efficient solution to guarantee reliability, transparency, and security concerning the data produced during the process. Our approach uses IoT technology for sensing and blockchain for storage. It is an excellent solution for applications related to the cold chain. Furthermore, our experiments results confirm the BlockClodChain's requirements satisfiability.



\bibliographystyle{plain}
\bibliography{references}

\begin{thebibliography}{10}

\bibitem{BADIAMELIS2018170}
R.~Badia-Melis, U.~{Mc Carthy}, L.~Ruiz-Garcia, J.~Garcia-Hierro, and J.I.
  {Robla Villalba}.
\newblock New trends in cold chain monitoring applications - a review.
\newblock {\em Food Control}, 86:170--182, 2018.

\bibitem{bumblauskas2020blockchain}
Daniel Bumblauskas, Arti Mann, Brett Dugan, and Jacy Rittmer.
\newblock A blockchain use case in food distribution: Do you know where your
  food has been?
\newblock {\em International Journal of Information Management}, 52:102008,
  2020.

\bibitem{buterin2013ethereum}
Vitalik Buterin et~al.
\newblock Ethereum white paper.
\newblock {\em GitHub repository}, pages 22--23, 2013.

\bibitem{de2020analysis}
Jos{\'e}~Eduardo de~Azevedo~Sousa, Vin{\'\i}cius Oliveira, J{\'u}lia Valadares,
  Glauber Dias~Gon{\c{c}}alves, Saulo Moraes~Villela, Heder Soares~Bernardino,
  and Alex Borges~Vieira.
\newblock An analysis of the fees and pending time correlation in ethereum.
\newblock {\em International Journal of Network Management}, page e2113, 2020.

\bibitem{figorilli2018blockchain}
Simone Figorilli, Francesca Antonucci, Corrado Costa, Federico Pallottino,
  Luciano Raso, Marco Castiglione, Edoardo Pinci, Davide Del~Vecchio, Giacomo
  Colle, Andrea~Rosario Proto, et~al.
\newblock A blockchain implementation prototype for the electronic open source
  traceability of wood along the whole supply chain.
\newblock {\em Sensors}, 18(9):3133, 2018.

\bibitem{francisco2018supply}
Kristoffer Francisco and David Swanson.
\newblock The supply chain has no clothes: Technology adoption of blockchain
  for supply chain transparency.
\newblock {\em Logistics}, 2(1):2, 2018.

\bibitem{GORDON2018224}
William~J. Gordon and Christian Catalini.
\newblock Blockchain technology for healthcare: Facilitating the transition to
  patient-driven interoperability.
\newblock {\em Computational and Structural Biotechnology Journal}, 16:224 --
  230, 2018.

\bibitem{hasanat2020iot}
Raisa~Tahseen Hasanat, MD~Arifur Rahman, Nafees Mansoor, Nabeel Mohammed,
  Mohammad~Shahriar Rahman, and Mirza Rasheduzzaman.
\newblock An iot based real-time data-centric monitoring system for vaccine
  cold chain.
\newblock In {\em 2020 IEEE East-West Design \& Test Symposium (EWDTS)}, pages
  1--5. IEEE, 2020.

\bibitem{kamble2020modeling}
Sachin~S Kamble, Angappa Gunasekaran, and Rohit Sharma.
\newblock Modeling the blockchain enabled traceability in agriculture supply
  chain.
\newblock {\em International Journal of Information Management}, 52:101967,
  2020.

\bibitem{kamilaris2019rise}
Andreas Kamilaris, Agusti Fonts, and Francesc~X Prenafeta-Bold$\acute{\nu}$.
\newblock The rise of blockchain technology in agriculture and food supply
  chains.
\newblock {\em Trends in Food Science \& Technology}, 91:640--652, 2019.

\bibitem{kshetri20181}
Nir Kshetri.
\newblock 1 blockchain’s roles in meeting key supply chain management
  objectives.
\newblock {\em International Journal of Information Management}, 39:80--89,
  2018.

\bibitem{WHOColdChain}
World~Health Organization.
\newblock The vaccine cold chain, 2015.

\bibitem{Vivaldi2020}
F.~Vivaldi, B.~Melai, A.~Bonini, N.~Poma, P.~Salvo, A.~Kirchhain, S.~Tintori,
  A.~Bigongiari, F.~Bertuccelli, G.~Isola, and F.~{Di Francesco}.
\newblock {A temperature-sensitive RFID tag for the identification of cold
  chain failures}.
\newblock {\em Sensors and Actuators, A: Physical}, 313:112182, 2020.

\bibitem{xu2019designing}
Xiwei Xu, Qinghua Lu, Yue Liu, Liming Zhu, Haonan Yao, and Athanasios~V
  Vasilakos.
\newblock Designing blockchain-based applications a case study for imported
  product traceability.
\newblock {\em Future Generation Computer Systems}, 92:399--406, 2019.

\end{thebibliography}

\end{document}